\documentclass[sigconf, nonacm]{acmart}
\usepackage{algorithm}
\usepackage{algpseudocode}
\usepackage{tabularx}
\usepackage{xcolor}
\usepackage{bbold}
\usepackage{enumitem}

\newcommand\vldbdoi{XX.XX/XXX.XX}
\newcommand\vldbpages{XXX-XXX}
\newcommand\vldbvolume{14}
\newcommand\vldbissue{1}
\newcommand\vldbyear{2020}
\newcommand\vldbauthors{\authors}
\newcommand\vldbtitle{\shorttitle} 
\newcommand\vldbavailabilityurl{URL_TO_YOUR_ARTIFACTS}
\newcommand\vldbpagestyle{plain}

\begin{document}
\title{LLM-R\textsuperscript{2}: A Large Language Model Enhanced Rule-based Rewrite System for Boosting Query Efficiency}

\settopmatter{authorsperrow=3}
\author{Zhaodonghui Li*}
\thanks{*: Zhaodonghui Li is under the Joint PhD Program between DAMO Academy and Nanyang Technological University}
\affiliation{%
  \institution{Nanyang Technological University, DAMO Academy Alibaba Group, Singapore}
}
\email{G220002@e.ntu.edu.sg}

\author{Haitao Yuan}
\affiliation{%
  \institution{Nanyang Technological University, Singapore}
}
\email{haitao.yuan@ntu.edu.sg}

\author{Huiming Wang**}
\thanks{**: Work done when interned in DAMO Academy}
\affiliation{%
  \institution{Singapore University of Technology and Design, Singapore}
}
\email{huiming_wang@mymail.sutd.edu.sg}

\author{Gao Cong}
\affiliation{%
  \institution{Nanyang Technological University, Singapore}
}
\email{gaocong@ntu.edu.sg}

\author{Lidong Bing}
\affiliation{%
  \institution{DAMO Academy, Alibaba Group, Singapore}
}
\email{l.bing@alibaba-inc.com}

\begin{abstract}
Query rewrite, which aims to generate more efficient queries by altering a SQL query’s structure without changing the query result, has been an important research problem. In order to maintain equivalence between the rewritten query and the original one during rewriting, traditional query rewrite methods always rewrite the queries following certain rewrite rules. However, some problems still remain. Firstly, existing methods of finding the optimal choice or sequence of rewrite rules are still limited and the process always costs a lot of resources. Methods involving discovering new rewrite rules typically require complicated proofs of structural logic or extensive user interactions. Secondly, current query rewrite methods usually rely highly on DBMS cost estimators which are often not accurate. In this paper, we address these problems by proposing a novel method of query rewrite named LLM-R\textsuperscript{2}, adopting a large language model (LLM) to propose possible rewrite rules for a database rewrite system. To further improve the inference ability of LLM in recommending rewrite rules, we train a contrastive model by curriculum to learn query representations and select effective query demonstrations for the LLM. Experimental results have shown that our method can significantly improve the query execution efficiency and outperform the baseline methods. In addition, our method enjoys high robustness across different datasets.
\end{abstract}

\maketitle

\pagestyle{\vldbpagestyle}
\begingroup\small\noindent\raggedright\textbf{PVLDB Reference Format:}\\
\vldbauthors. \vldbtitle. PVLDB, \vldbvolume(\vldbissue): \vldbpages, \vldbyear.\\
\href{https://doi.org/\vldbdoi}{doi:\vldbdoi}
\endgroup
\begingroup
\renewcommand\thefootnote{}\footnote{\noindent
This work is licensed under the Creative Commons BY-NC-ND 4.0 International License. Visit \url{https://creativecommons.org/licenses/by-nc-nd/4.0/} to view a copy of this license. For any use beyond those covered by this license, obtain permission by emailing \href{mailto:info@vldb.org}{info@vldb.org}. Copyright is held by the owner/author(s). Publication rights licensed to the VLDB Endowment. \\
\raggedright Proceedings of the VLDB Endowment, Vol. \vldbvolume, No. \vldbissue\ %
ISSN 2150-8097. \\
\href{https://doi.org/\vldbdoi}{doi:\vldbdoi} \\
}\addtocounter{footnote}{-1}\endgroup


\ifdefempty{\vldbavailabilityurl}{}{
\vspace{.3cm}
\begingroup\small\noindent\raggedright\textbf{PVLDB Artifact Availability:}\\
The source code, data, and/or other artifacts have been made available at \url{https://github.com/DAMO-NLP-SG/LLM-R2}.
\endgroup
}


\section{Introduction}
With the rapid growth of data in various fields, it is common to take seconds or minutes, and even longer to execute an SQL query. Therefore, efficient query processing has been a crucial task in modern database systems. One of the key topics in query optimization that has gained significant attention is query rewrite \cite{10.1145/141484.130294, 10.14778/3352063.3352141}. The objective of query rewrite is to output a new query equivalent to the original SQL query, while having a shorter execution time. Ideally, query rewrite should fulfill three critical criteria: (1) \textbf{Executability}: the rewritten query should be able to be executed without any errors; (2) \textbf{Equivalence}: it must yield identical results as the original query; (3) \textbf{Efficiency}: this encompasses two aspects—\textit{Execution Efficiency} and \textit{Computational Efficiency}. \textit{Execution Efficiency} refers to the requirement that the rewritten query executes more efficiently than the original, while \textit{Computational Efficiency} implies that the overhead of the rewriting process should be justifiable by the time savings achieved during query execution.

To enhance both \textbf{Executability} and \textbf{Equivalence} in rewritten queries, existing studies have predominantly concentrated on rule-based rewriting techniques. In particular, these studies are divided into two orthogonal research directions: the discovery of novel rewriting rules and the effective application of existing rules. For the first direction, although some studies have discovered more rewrite rules~\cite{10.1145/3514221.3526125, bai2023querybooster, wu2022factor}, there are many challenges related to the complexity of rule validation and the specificity of their applicability, often resulting in high computational demands and professional-level user competence. For example, Wetune~\cite{10.1145/3514221.3526125} only supports discovering rewrite rules on limited types of operators and Querybooster~\cite{bai2023querybooster} necessitates user engagement with specialized rule syntax, respectively. Therefore, this paper shifts focus toward the latter direction, delving into the methodologies for the effective utilization of pre-established rules. For example, Learned Rewrite \cite{10.14778/3485450.3485456} utilizes existing rewrite rules from the Apache Calcite \cite{Begoli_2018} platform and learns to select rules to apply. It notably incorporates a Monte Carlo search algorithm in collaboration with a machine-learned query cost estimator to streamline the selection process. However, it's non-trivial to solve the challenges related to the computational demand of the Monte Carlo algorithm and the precision of the cost estimation model, which can significantly impact the \textbf{execution efficiency}.

On the other hand, with the rise of large language models (LLMs), there also exist some ``large language model for database'' projects \cite{xue2024dbgpt, db-gpt-qh} that support direct query rewrite. The idea of these methods is to utilize the sequence-to-sequence generation ability of a language model to directly output a new rewritten query given an input query, without considering any rewrite rules or DBMS information. Although it is possible for these methods to discover new rewrites not following any existing rules, they easily suffer from the hallucination problem of language models \cite{Ji_2023, zhang2023sirens}, especially for long and complicated queries, where language models give plausible but incorrect outputs. Either a syntax or reference error during generation will lead to vital errors when executing the query. Therefore, relying solely on LLM’s output query may violate the \textbf{executability} and \textbf{equivalence} to the original query, deviating from the basic aim for query rewrite.

To overcome the limits of the current query rewriting techniques and benefit from their advantages, we propose an LLM-enhanced rewrite system to use LLMs to suggest rewrite rule strategies and apply these strategies with an existing database platform to rewrite an input query. Inspired by the LLM-based learning framework for using tools \cite{schick2023toolformer, yao2023react}, we leverage the LLM's strong generalization and reasoning abilities for query rewriting while avoiding issues like hallucination. We design a novel LLM-enhanced query rewrite system to automate the process of selecting more effective rewrite rules, note that the \textbf{executability} and \textbf{equivalence} of the rewritten query are guaranteed since all the candidate rules are provided by existing DB-based rule rewrite platforms. In addition to meeting the basic requirements of valid query rewrite, we also develop new techniques to boost the \textbf{executing efficiency} of our rewrite system. Firstly, to overcome hallucination, we collect a pool of demonstrations consisting of effective query rewrites using existing methods and our designed baselines. We then learn a contrastive query representation model to select the most useful in-context demonstration for the given query to prompt the system, optimizing the LLM's rewrite rule selection. In addition, to address the challenge of limited training data, we propose to utilize the learning curriculum technique \cite{10.1145/1553374.1553380} to schedule the training data from easy to hard. We apply our LLM-enhanced rewrite method on three different datasets, namely TPC-H, IMDB, and DSB. We observe a significant query execution time decrease using our method, taking only 52.5\%, 56.0\%, 39.8\% of the querying time of the original query and 94.5\%, 63.1\%, 40.7\% of the time of the state-of-the-art baseline method on average on the three datasets.

Our main contributions are:
\begin{itemize}[leftmargin=10.2pt]
\vspace{-\topsep}
\setlength{\itemsep}{0pt}
\setlength{\parsep}{0pt}
\setlength{\parskip}{0pt}
  \item To the best of our knowledge, we are the first to propose an LLM-enhanced query rewrite system that can automatically select effective rules from a given set of rewrite rules to rewrite an input SQL query.
  \item To enable LLMs to select better rewrite rules for a query, we construct a demonstration pool that contains high-quality demonstrations so that we can select good demonstrations to prompt the LLM-enhanced rewrite system for few-shots learning. 
  \item We learn a contrastive query representation model to optimize the demonstration selection. To overcome the challenge of limited training data, we further design a learning curriculum to schedule the training data from easy to hard.
  \item We further analyze the robustness of our method. By applying our method to unseen datasets and different dataset volumes, we demonstrate that our method is much more flexible than the baseline methods and shed light on generalizing to other database problems.
\end{itemize}

\section{Preliminary}

\begin{table*}
\newcommand{\wrap}[1]{\parbox{.67\linewidth}{\vspace{1.5mm}#1\vspace{1mm}}}
\caption{Examples of query rewrite rules. Examples of query rewrite rules of the Apache Calcite Rules \cite{org.apache.calcite.rel.rules}.}
\label{tab:rules}
\begin{tabular}{|l|l|} \hline
\textbf{Rule Name} & \textbf{Rule Description}  \\ \hline
AGGREGATE\_UNION\_AGGREGATE                  & Rule that matches an Aggregate whose input is a Union one of whose inputs is an Aggregate \\ \hline
FILTER\_INTO\_JOIN                                & Rule that tries to push filter expressions into a join condition and into the inputs of the join  \\ \hline
JOIN\_EXTRACT\_FILTER                             & Rule to convert an inner join to a filter on top of a cartesian inner join  \\  \hline
SORT\_UNION\_TRANSPOSE                            & Rule that pushes a Sort past a Union  \\ \hline
\end{tabular}
\end{table*}

In this section, we first introduce some key concepts including query, query tree and query rewrite rules in Section \ref{sec:def_rewriterule}. Then, we will formalize the problem of query rewrite based on rules in Section~\ref{sec:def_qrewrite}. Finally in Section \ref{sec:def_relatedwork}, we introduce the related work.

\begin{figure}[t]
  \centering
  \includegraphics[width=0.9\linewidth]{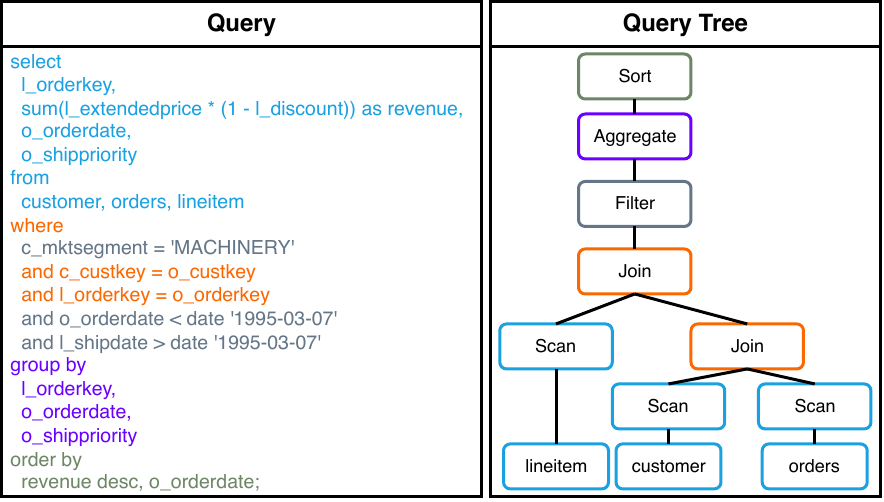}
  \vspace{-10pt}
  \caption{A TPC-H query and its query tree}
  \Description{plan_tree}
  \vspace{-15pt}
  \label{fig:plan_tree}
\end{figure}
\subsection{Query and Rewrite Rules}
\label{sec:def_rewriterule}
\textbf{Query \& Query tree.} Each query in our study is formulated as an executable SQL statement. Furthermore, we model each query as a query tree using various nodes, where each node represents a specific type of query operator (e.g., Sort, Join, and Scan). Figure~\ref{fig:plan_tree} illustrates an example of a SQL query and its corresponding query tree representation. It is worth noting that any given query can be transformed into a query tree, and conversely, this query tree can be reverted back to its original raw query form.

\noindent \textbf{Query rewrite rules.} Given an input query denoted as $Q$, a sequence of transformation methods, represented as $r_1, r_2, \cdots$, can be applied to the query's query tree, yielding an equivalent query, denoted as $Q^*$. These transformation methods, referred to as rewrite rules, encompass a diverse range of functionalities. These include the conversion of one operator to another, the alteration of execution sequences between operators, and the elimination of redundant operators. Table~\ref{tab:rules} delineates a representative set of these query rewrite rules. For the sake of brevity, we succinctly express the query rewrite process as $Q^* = R(Q)$, where $R = [r_1, r_2, \cdots, r_n]$ symbolizes the sequence of $n$ applied rewrite rules.


\subsection{Rule-based Query Rewrite}
\label{sec:def_qrewrite}

With the introduction of the rewrite rules, we now formally define the problem of query rewrite based on rules as follows:
\begin{definition}
\label{def:rewrite_objective}
(Rule-based query rewrite): Consider an input query $Q$ and a set of candidate rewrite rules $R$. The objective is to identify a sequence of rules $R^* = [r^*_1, r^*_2, \cdots, r^*_n]$ where $  r^*_i \in R$, that transforms the query $Q$ into a more efficient version $Q^* = R^*(Q)$. The efficiency of the rewritten query $Q^*$ is quantified by its execution latency. Such rewrite is characterized by transforming $Q$ into an equivalent query $Q^*$, which exhibits a lower execution latency compared to other possible rewritten versions of the query. The problem can be formally represented as:
\begin{equation}
\begin{aligned}
\text{argmin}_{R^* \subseteq R} & \text{ latency}(Q^*) \\
\text{s.t.}\ \  & Q^* = R^*(Q)
\end{aligned}
\end{equation}
\end{definition}

\subsection{Related Work}
\label{sec:def_relatedwork}


\subsubsection{Query Rewrite}

Query rewrite is a significant function in current Database Management Systems (DBMSs), and can be supported in the query optimizers \cite{344061, Graefe1995TheCF, 10.1145/38713.38734}. In particular, DBMSs, such as Calcite~\cite{Begoli_2018} and PostgreSQL~\cite{PostgreSQL}, have developed different rewrite functions to achieve various rewrite rules. Consequently, there are two primary research directions for the query rewriting problem: discovering new rewrite rules and optimally leveraging existing rewrite rules.

\noindent \textbf{Discovering New Rewrite Rules.} 
Recent advancements, exemplified by Querybooster~\cite{bai2023querybooster} and Wetune~\cite{10.1145/3514221.3526125}, have made significant strides in discovering new rewrite rules through the application of relational algebra proofs \cite{wu2022factor}. Querybooster enables database users to suggest rules through a specialized rule language, facilitating the back-end generation and application of these rules for more adaptable rewriting. On the other hand, Wetune compiles potential rewrite templates and pinpoints constraints that convert these templates into actionable rules. While these methodologies have proven their worth by efficiently handling small real-world workloads, they have their limitations. Querybooster's effectiveness hinges on the user's ability to propose potent rules, whereas Wetune's efficacy on simple or generalized queries remains uncertain.

\noindent \textbf{Selecting Rewrite Rules.} The heuristic rewrite approach executes rewrite rules contingent upon the types of operators involved. Nonetheless, this technique is not without flaws. It might not identify the most optimal sequences for rewriting and often lacks the mechanisms necessary for evaluating the benefits of such rewrites. To address this issue, Learned Rewrite~\cite{10.14778/3485450.3485456} employs a Monte Carlo Tree search to optimize the selection of applicable rules. It conceptualizes each query as a query tree, with applied rules modifying the tree's structure. This approach utilizes a learned cost model to predict the impact of applying specific rules, enabling the selection of an optimal rewrite sequence through Monte Carlo Tree search. While this method improves adaptability to varying queries and database structures, it faces challenges in cost model accuracy and potential local minima in the search process, highlighting areas for future enhancement in rule-based query rewriting techniques.

\subsubsection{LLM-based SQL Solvers.} 
Large Language Models (LLMs) have recently emerged as a hot topic in machine learning research, captivating the interest of many in the field due to their impressive capabilities. These models have demonstrated a surprisingly strong ability to handle a variety of text-related tasks, excelling in areas such as generation, decision-making, and deduction. One such task that is highly related to DB research is text-to-SQL, in which an LLM directly generates a SQL query given database information and user requirements. Numerous studies~\cite{li2023llm, sun2023sqlpalm, zhou2023r3nl2gql} have highlighted the potential of LLMs in the text-to-SQL task, showcasing their proficiency in SQL query-related tasks. While much of this existing research has focused on LLMs' ability to generate executable queries, there is a growing recognition of the importance of other factors, particularly the efficiency and accuracy of these queries when applied in real-world scenarios. In particular, \cite{li2023llm} discussed their attempts in an efficiency-oriented query rewrite task, where an LLM is directly given an input query and tries to rewrite it into a more efficient one.

However, a significant issue previous LLM-based face is the problem of hallucination, which refers to instances where the model generates output that is not only incorrect but is done so with a misleading level of confidence. This is particularly problematic in the context of database applications, where accuracy is paramount. Therefore, we propose a different direction of utilising the LLMs while overcoming hallucination. Instead of using LLM to directly output an SQL query, we adopted a DB-based SQL rewriter enhanced by an LLM.

\subsubsection{In-context Learning} Due to the extensive data and resource requirements of fine-tuning an LLM, many works choose to utilize LLMs by the technique called in-context learning (ICL), where no modifications to the LLMs' model weights are made. The concept of ICL, first introduced by \citeauthor{brown2020language} in their seminal work on GPT-3 \cite{brown2020language}, shows that language models like GPT-3 can leverage in-context demonstrations at inference time to perform specific tasks, without updating the model weights. ICL typically involves enriching the context with select examples to steer the model's output. Formally, consider a model denoted as $M$ and a contextual input represented by $P$. The output $o$ generated by applying the ICL method to model $M$ with input $P$ can be succinctly expressed as $o = ICL_M(P)$.

ICL has rapidly gained popularity for addressing diverse challenges in natural language processing. However, it is a sophisticated technique requiring careful implementation. Extensive research, including studies by \cite{wei2023larger} and \cite{li2023unified}, has explored the intricacies of LLMs' learning processes in this context. These studies highlight that the success of in-context learning is closely related to the construction of the context and the quality of the examples used.

\section{LLM-enhanced Rewrite System}
In this section, we will introduce our innovative LLM-enhanced rule-based rewrite system (\textbf{LLM-R\textsuperscript{2}}). In Section \ref{sec:system_pipeline}, we will first illustrate the pipeline of our rewrite system. Then in Section \ref{sec:demonstration_selection}, we will state our motivation to optimize the demonstration selection and introduce our novel \textit{Demonstration Manager} module.

\begin{figure}[t]
  \centering
  \includegraphics[width=\linewidth]{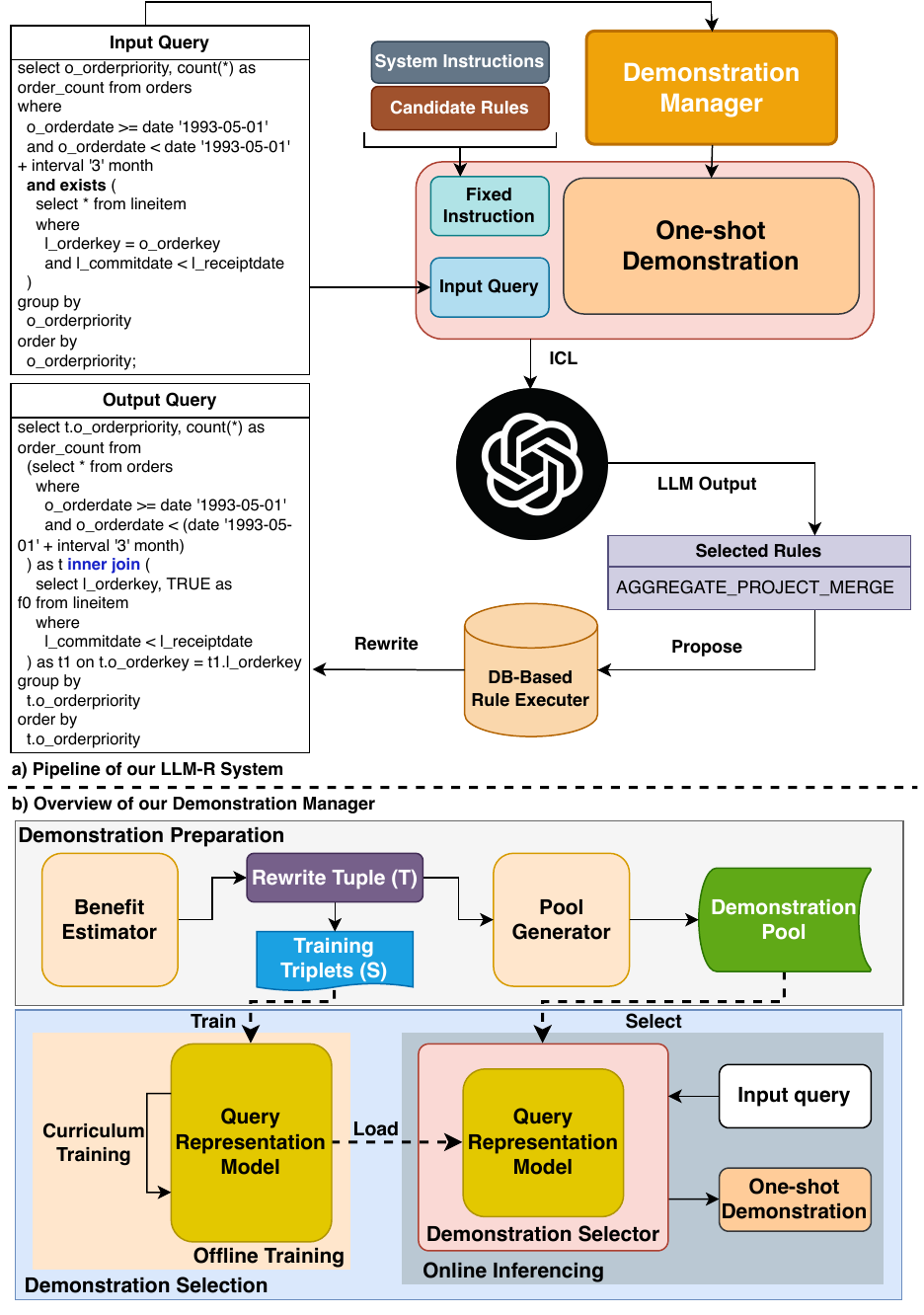}
  \vspace{-15pt}
  \caption{The Framework of LLM-enhanced Rewrite System}
  \vspace{-10pt}
  \Description{LLM-R}
  \label{fig:LLM-R}
\end{figure}

\subsection{System Pipeline}
\label{sec:system_pipeline}

As shown in Figure ~\ref{fig:LLM-R}(a), the system integrates an LLM into the query rewrite system utilizing the ICL methodology \cite{brown2020language}. We construct the ICL prompt with three main components:

\noindent \textbf{Input query:} We employ the SQL statement corresponding to the provided input query $Q$ for the prompt construction.

\noindent \textbf{Fixed instruction:} The fixed instruction consists of a system instruction $I$ and a rule instruction $R$. While the system instruction specifies the task requirements, the rule instruction includes a comprehensive list of all candidate rewrite rules available for the language model to select. Each rule is accompanied by a concise explanation, enabling informed decision-making.


\noindent \textbf{One-shot demonstration:} Similar to directly using LLMs to rewrite queries, selecting rewrite rules using LLMs may also easily suffer from the hallucination problem, like outputting non-existing rules. To mitigate this and ensure the LLMs' outputs are more closely aligned with our task requirements, yielding superior rule suggestions, we use the demonstration as a part of the prompt. Formally, we define our demonstration given to the LLM-R\textsuperscript{2} system as a pair of text $D=\langle Q^D, R^D\rangle$, where $Q^D$ is the example query assembling the input query and $R^D=[r^D_1,\cdots]$ is the list of rules that have been applied to rewrite the example query. Such demonstrations can successfully instruct the LLM to follow the example and output a list of rewrite rules to apply on the new input query. In particular, this involves selecting a high-quality demonstration $D$ from many successful rewritten demonstrations (i.e., denoted as a pool $\mathcal{D}$) for each input query to guide the LLM effectively. To achieve this goal, we design a module named \textit{Demonstration Manager}, whose details are elucidated in the subsequent section.

\begin{figure*}[t]
  \centering
  \includegraphics[width=\linewidth]{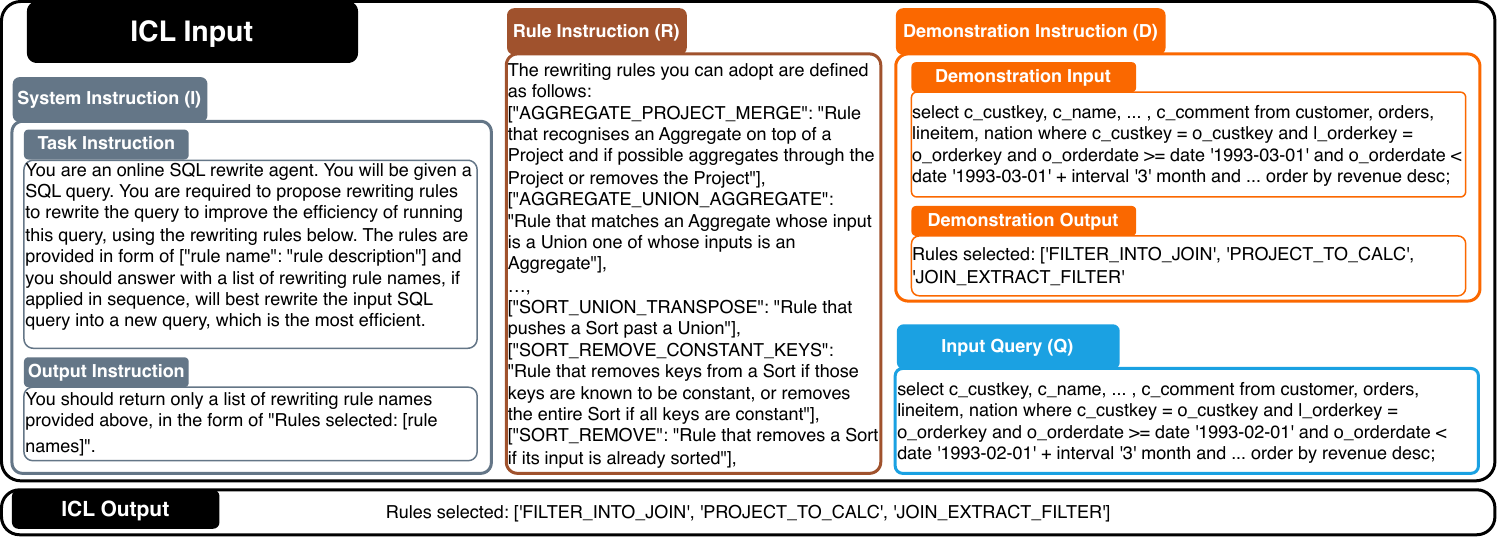}
  \vspace{-15pt}
  \caption{An Example of the In-Context Learning Process in LLM-R\textsuperscript{2}. All the instructions are concatenated together as one string input to the LLM. In a zero-shot setting, the ``Demonstration Instruction'' will be removed and an input query will be appended directly after the ``Rule Instruction''.}
  \vspace{-10pt}
  \Description{icl_example}
  \label{fig:icl_example}
\end{figure*}

As specifically highlighted, Figure~\ref{fig:icl_example} delineates the prompt utilized within the In-Context Learning (ICL) process of our system. Upon constructing the prompt and feeding it into the LLM, we can extract a sequence of rewrite rules from the model's output. These rules undergo further processing and execution by a database-based rule executor. For instance, the original input query in Figure \ref{fig:LLM-R}(a) is modified by the ``AGGREGATE\_PROJECT\_MERGE'' rule, as highlighted in bold. This modification transforms the original query into a more optimized output query, demonstrating the practical application and effectiveness of the extracted rules in query optimization processes. Through the synergy of the LLM's superior generalization capabilities and the rule executor's precision, our proposed system guarantees extensive applicability, alongside ensuring the executability and equivalence of the rewritten queries. Consequently, this rewrite process can be formalized as follows:

\begin{definition}
(LLM-enhanced Query Rewrite): Given a large language model $M$, a textual instruction outlining the rewrite task $I$, a set of candidate rules $R$, one successful rewrite demonstration $D$ selected from the demonstration pool $\mathcal{D}$, and an input query $Q$, a prompt $P$ is constructed and provided as input to $M$ as:
\begin{displaymath}
P = I \oplus R \oplus D \oplus Q
\end{displaymath}
From $M$, a sequence of rewrite rules $R^*$ is derived:
\begin{displaymath}
R^* = ICL_M(P)
\end{displaymath}
By sequentially applying these rewrite rules $R^*$, we generate an optimally equivalent query, represented as $Q^* = R^*(Q)$. 
\end{definition}


\begin{figure}[t]
  \centering
  \includegraphics[width=\linewidth]{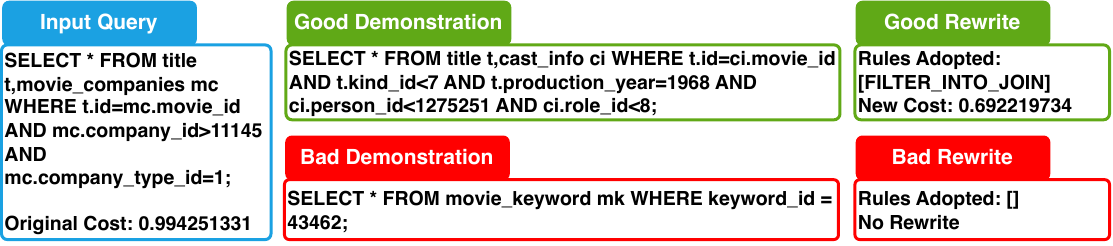}
  \vspace{-15pt}
  \caption{Example of good and bad demonstration selections}
  \vspace{-10pt}
  \Description{demonstration_example}
  \label{fig:demonstration_example}
\end{figure}

\subsection{Demonstration Manager Overview}
\label{sec:demonstration_selection}

\noindent \textbf{Motivation.} In the above ICL process, optimizing the prompt $P=I \oplus R \oplus D \oplus Q$ is crucial for improving the output quality of LLMs. Given the fixed settings of system instruction($I$), rule instruction($R$), and input query($Q$), our optimization efforts focus primarily on the demonstration($D$), which is chosen to enhance model performance. Recent studies on LLMs (e.g., ~\cite{brown2020language,wei2023larger}) have underscored the positive impact of high-quality in-context demonstrations on LLM output, reducing the tendency of LLMs to produce hallucinatory content. As shown in Figure~\ref{fig:demonstration_example}, our rewrite system exhibits similar effectiveness variability w.r.t. the demonstrations used, further emphasizing the necessity of optimizing demonstration selection for specific input queries. Therefore, it is an important problem to optimize the demonstration selected for a given input query. Particularly, we address this problem by designing the \textit{Demonstration Manager} module.

\noindent \textbf{Overview.} Figure \ref{fig:LLM-R}(b) illustrates the basic structure of our proposed \textit{Demonstration Manager} module, comprising two  parts: \textit{Demonstration Preparation} and \textit{Demonstration Selection}. 

\noindent (1) The primary objective of the Demonstration Preparation is to generate a substantial number of successful rewritten demonstrations for constructing a demonstration pool. Furthermore, this part also serves to supply training data essential for model learning in the second part. Specifically, we design two modules: the \textit{Benefit Estimator} and the \textit{Pool Generator}, to achieve our objectives. The \textit{Benefit Estimator} is capable of assessing the potential benefits of a given query rewrite strategy, thereby generating corresponding rewrite tuple recording the performance of this rewrite strategy on the input query. Subsequently, the \textit{Pool Generator} is employed to extract demonstrations for constructing a pool. Moreover, we utilize the rewrite tuples to derive training triplets, which are essential for model learning in subsequent parts.

\noindent (2) The second part involves the \textit{Demonstration Selection} module, tasked with identifying the optimal demonstration from the pool for each input query. This process is enhanced by incorporating a query representation model within the selector, designed to evaluate the similarity between input queries and demonstrations in the pool. This representation model undergoes offline training using the training data. In addition, to obtain an effective model, we enhance the model’s training through the integration of a curriculum learning approach. Afterwards, the trained model is integrated into \textit{Demonstration Selector} for online inference. In other words, upon receiving an input query for rewriting, the selector discerns and selects the most appropriate demonstration from the pool based on the trained model. More detailed elaboration on the above two parts will be provided in the following sections.

\section{Demonstration Preparation}
\label{sec:method_DM}

\begin{figure}[t]
  \centering
  \includegraphics[width=\linewidth]{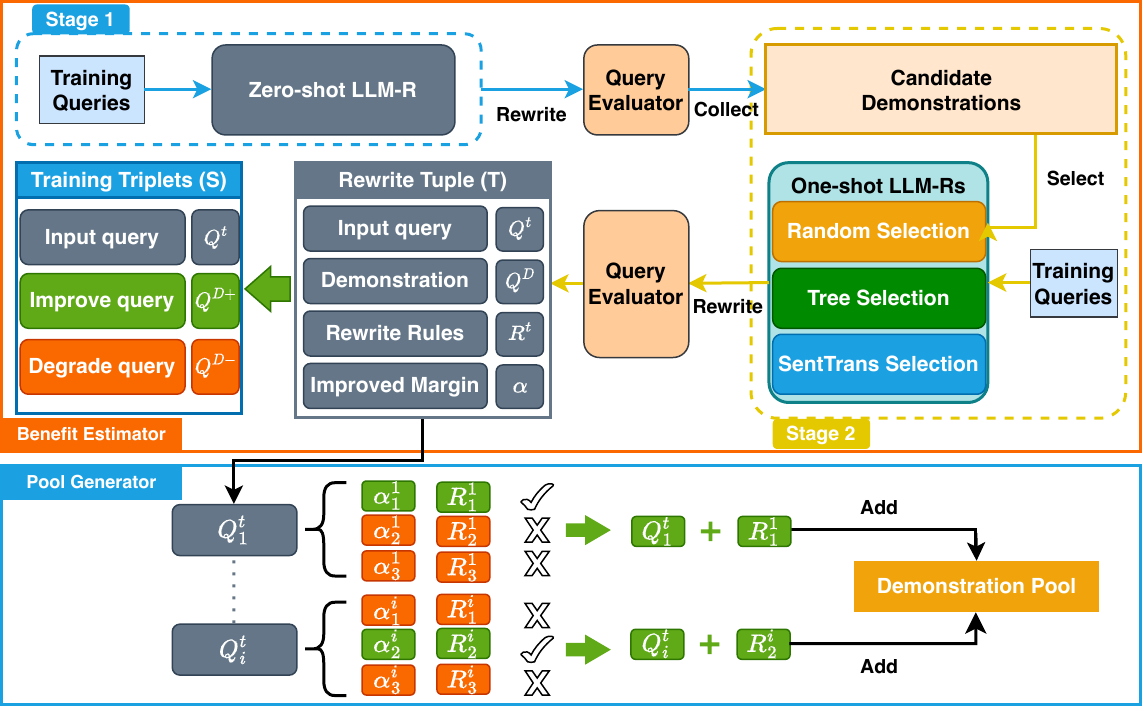}
  \vspace{-15pt}
  \caption{Our demonstration preparation module generates a set of training triplets and a demonstration pool.}
  \vspace{-10pt}
  \Description{data manager}
  \label{fig:DM}
\end{figure}

In this section, we aim to generate sufficient high-quality data to build the demonstration pool. As shown in Figure~\ref{fig:DM}, we first design the \textit{Benefit Estimator} module to generate the ground truth, where each ground truth data point indicates the efficiency gain obtained by rewriting an input query using generated rules in the context of a demonstration. With sufficient ground truth, including both good and bad samples, we further design the \textit{Pool Generator} module to select all good samples to build the demonstration pool. In addition, we can deduce contrastive training triplets from the ground truth, which can help train our selection model.

\subsection{Benefit Estimator}
Since we are only able to start with solely training queries without demonstrations, the triplet generation pipeline is segmented into two distinct phases: the first stage involves initializing high quality candidate demonstrations utilizing baseline method and a zero-shot \textbf{LLM-R\textsuperscript{2}} system where no demonstration is selected, followed by the demonstration adoption stage employing a one-shot \textbf{LLM-R\textsuperscript{2}} system. Subsequently, each stage is elucidated in detail.

\noindent \textbf{Stage-1:} We start with a diverse set of input queries collected from our dataset as the training set. To obtain a rich set of effective rewrites as candidate demonstrations, we first apply our zero-shot LLM-enhanced rewrite system (LLM-R\textsuperscript{2}) to rewrite the training set queries. After getting the rewrite rules adopted and the resulted rewrite queries, we directly execute the rewritten queries on the corresponding databases. The execution time of the rewritten queries as well as the original queries is evaluated to collect the initial candidate demonstration set consisting of the improvable queries, together with their rules adopted.

\noindent \textbf{Stage-2:} With the candidate demonstrations collected from the previous step, we can then estimate the benefits of these demonstrations when they are selected for a given input query. Motivated by \cite{wei2023larger}, such improvable demonstrations are supposed to be more useful for the LLM to output improving rewrite suggestions, compared to using any degraded rewrite queries as demonstrations. In addition, the more ``similar'' the improving demonstration query is to the input query, the better output the LLM will generate. However, different from natural language inputs’ simple textual similarity, the similarity between SQL queries is indeed more complicated. To identify if the pool we collected truly contains high-quality and ``similar'' demonstrations for new input queries and refine the demonstration pool, we designed three heuristic demonstration-selection methods based on different levels of similarity as follows. 



\vspace{-\topsep}
\begin{itemize}[leftmargin=10.2pt]
\setlength{\itemsep}{0pt}
\setlength{\parsep}{0pt}
\setlength{\parskip}{0pt}
\item  \textbf{Random Selection:} A random demonstration query is selected from the candidate demonstrations for a given input query, where the similarity level lies on the same input category.

\item \textbf{Tree Selection:} Query tree is an important structural feature for the queries, therefore, it is natural to align similarity with the query tree structure. We first compute the query trees of all the candidate demonstration queries, with operators as the tree nodes. Given an input query, we select the demonstration with the minimum tree edit distance from the input query tree within the candidate demonstrations.

\item \textbf{SentTrans Selection:} At the textual level, we observe that queries are always considered as sentences for the language models to process. Based on the observation, we treat input queries as sentences and select the candidate demonstration query whose embedding is the most similar to the input query. Most of the effective LLMs are closed-sourced, which means we are not able to obtain the query embeddings of such LLMs. However, similar to LLMs, some small pre-trained language models share the same sequence-to-sequence mechanism, that the input text is first encoded by an encoder before feeding to the model. Using such encoders, like Sentence Transformers \cite{reimers-2019-sentence-bert}, we can obtain an embedding of a given sentence. 
\end{itemize}
\vspace{-\topsep}
With the three demonstration selection methods above, we can prompt our \textbf{LLM-R\textsuperscript{2}} system with the one-shot demonstration to obtain various rewrite results on the same training set. These new rewrite queries from the one-shot \textbf{LLM-R\textsuperscript{2}} system are then evaluated in the same way as in Stage-1. Specifically, when we adopt one-shot demonstration to rewrite an input query $Q^t$, we are able to estimate the benefit obtained from the demonstration by constructing the rewrite tuples (T) as ($Q^t, D, R^t, \alpha$), where $Q^t$ represents a training query, $D$ is the demonstration $\langle Q^D, R^D \rangle$ selected for $Q^t$, $R^t$ denotes the adopted rules for $Q^t$, and $\alpha$ represents the improved margin obtained by the query rewrite. In particular, given the original query cost $C_0$ and the cost of rewritten query $C_r$, we define the improved margin as $\alpha=C_0/C_r$, where the larger margin the better rewrite result and larger benefit we have.

In addition, a set of training triplets is generated using the rewrite tuples obtained in preparation for training a contrastive representation model. For a given query $Q^t$ in the rewrite tuple ($Q^t, D, R^t, \alpha$), we consider the demonstration query $Q^D$ adopted as an improve query $Q^{D+}$ for $Q$, if the improved margin $\alpha>1$. In contrast, we denote the demonstration query as a degrade query $Q^{D-}$ if $\alpha<1$. If there are multiple improve(degrade) queries, we only select the one with the largest(smallest) improved margin. Since we have adopted multiple one-shot selection methods, now we are able to construct a training triplet for a given query as $\langle Q^t, Q^{D+}, Q^{D-}\rangle$. A set of training triplets can be further constructed if we enumerate the whole training query set.

\subsection{Pool Generator}

Apart from the training triplets, we also hope to prepare an effective demonstration pool so that our learned demonstration selection model can select demonstrations from it during online inference. The rewrite tuple generated by the \textit{Benefit Estimator} module, recording the effectiveness of a sequence of rewrite rules $R^t$ on an input query $Q^t$, naturally fits our need for a high-quality rewrite demonstration. 

In particular, given the set of rewrite tuples generated by $n$ input queries, we first separate them into $n$ groups $\{T_i\}_{1 \leq i \leq n}$ based on their corresponding input queries. Therefore, each group $T_i$ can be represented as the tuple set $T_i=\{(Q_i^t, D^i_1, R_1^i, \alpha_1^i), (Q_i^t, D_2^i, R_2^i, \alpha_2^i), \cdots\}$. Since we have adopted various methods, multiple tuples have the same input query, and we only need the optimal rewrite rule sequence to form a demonstration for the query. Therefore, for each training query $Q_i^t$ and its corresponding tuple group $T_i$, we only select the tuple with the largest improved margin, and the order is denoted as $*$, which can be formulated as follows:
\begin{equation}
\begin{aligned}
* &= argmax_{j \in [1, |T_i|]} \alpha_{j}^i\\
s.t. &  T_i=\{(Q_i^t, D^i_1, R_1^i, \alpha_1^i), (Q_i^t, D_2^i, R_2^i, \alpha_2^i), \cdots\}
\end{aligned}
\end{equation}
Next, we construct the demonstration containing the input query and rules as the pair $\langle Q^t_i, R_*^i \rangle$, and then add the demonstration to the pool.
As shown in Figure~\ref{fig:DM}, when the largest improved margins $\alpha_1^1$ and $\alpha_2^i$ are identified for input queries $Q_1^t$ and $Q_i^t$, the corresponding demonstrations $\langle Q_1^t, R_1^1 \rangle$ and $\langle Q_i^t, R_2^i \rangle$ are selected with the rewrite rules $R_1^1$ and $R_2^i$ adopted.



\begin{figure}[!t]
  \centering
  \includegraphics[width=\linewidth]{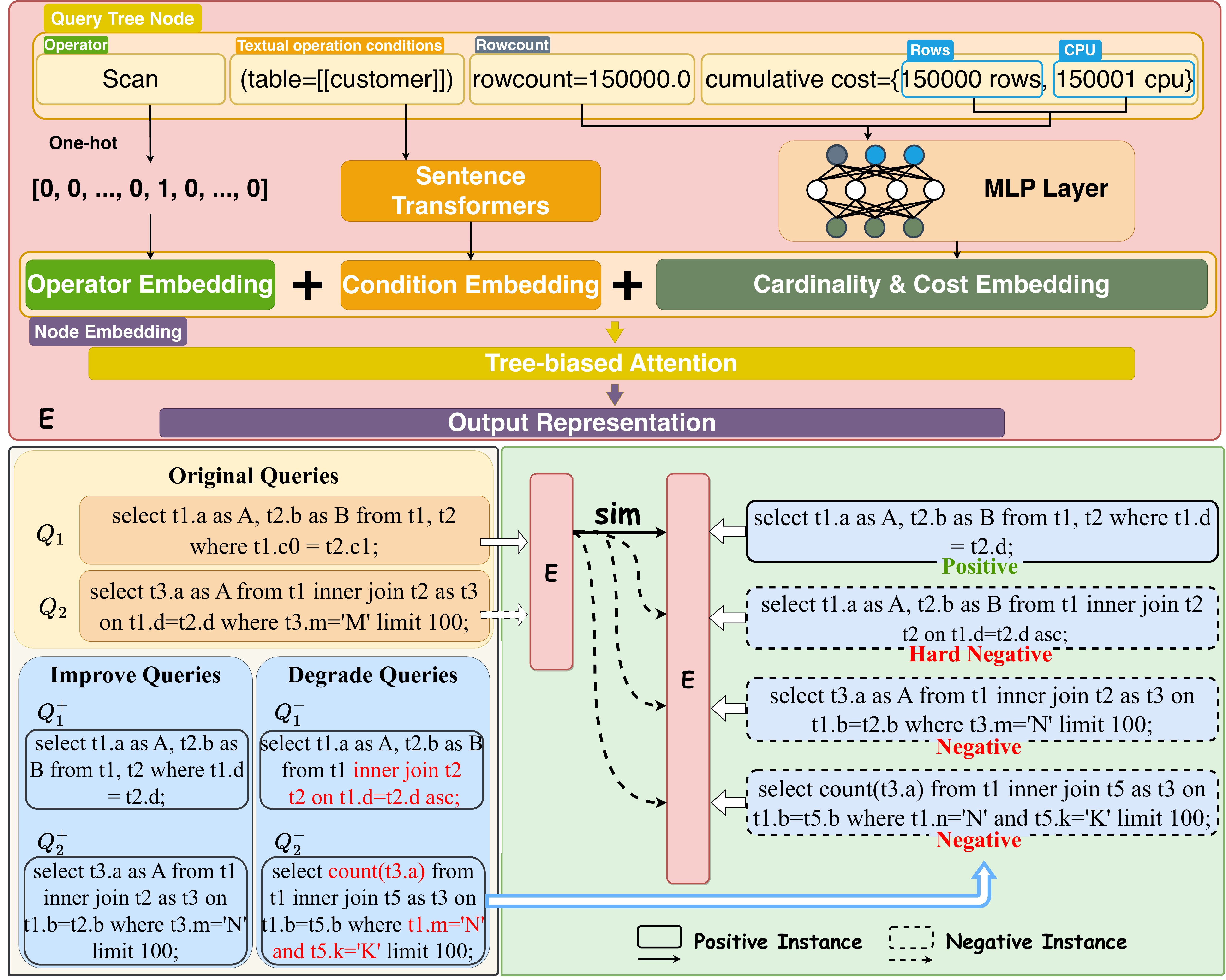}
  \vspace{-15pt}
  \caption{Our representation model and its contrastive training structure. Each node of the query tree is encoded into a fixed length vector, while the final representation of the query is obtained by applying a tree-biased attention over the tree nodes. Such model $E$ is then trained with contrastive query tuples generated.}
  \vspace{-15pt}
  \Description{demonstration selector}
  \label{fig:CS}
\end{figure}

\section{Demonstration Selection}
\label{sec:method_CS}

\textbf{Motivation.} Addressing the challenge of enhancing system performance, the selection of an optimal rewrite demonstration to guide the LLM for any given input query is required and remains uncertain. Intuitively, the greater the ``similarity'' between the input and demonstration queries, the more applicable the rewrite rule, thereby enhancing the LLM's output efficacy. Therefore, to capture such ``similarity'', we design a contrastive model to learn the representations of queries in this \textit{Demonstration Selection} module, where better demonstration queries are to have more similar representations to the input query. Consequently, the demonstration query that exhibits the highest resemblance to the input query is selected for the LLM, optimizing the generation of more effective outputs. 

\noindent\textbf{Overview.} In order to learn a contrastive representation model efficiently and effectively, the selection module consists of two main components: our contrastive model and a curriculum learning pipeline to improve the model training. We will first outline the representation model and its contrastive learning structure in Section~\ref{sec:mod_str}, followed by a detailed discussion of the whole model learning pipeline in Section~\ref{sec:mod_lea}.

\subsection{Contrastive Representation Model\label{sec:mod_str}} 
As shown in Figure \ref{fig:CS}, our representation model $E$ is constructed as a query encoder to encode the information describing a query, and a contrastive training structure to further train the encoder given training data. In particular, the information of a query tree is first encoded by nodes into node embeddings. A tree-biased attention layer will then compute the final representation of the query given the node embeddings. Such an encoder $E$ is then trained using the contrastive learning structure drawn below it.

\noindent \textbf{Query encoder.} The representation of a query should focus on various key attributes, like the query tree structure and columns selected. Therefore, we design an encoder following \cite{article} to take the query trees generated by DBMS' query analyzer as inputs. It is notable that the original encoding in \cite{article} utilizes the physical query plan which contains richer information, so that the objective of estimating query cost can be successfully achieved. Since we aim to capture the similarity between queries, we separately encode the following information for each query tree node instead in our encoder, as shown in the top half of Figure \ref{fig:CS}:

\vspace{-\topsep}
\begin{itemize}[leftmargin=10.2pt]
\setlength{\itemsep}{0pt}
\setlength{\parsep}{0pt}
\setlength{\parskip}{0pt}
  \item \textbf{Operator type:} We use one-hot encoding to encode the operator types into one vector, with value one for the current node operator type and zero for the rest positions.
  \item \textbf{Operator conditions:} Within each node, the details for the operator are explained in parentheses, including sort order for “Sort” operator, selected column for “Scan” operator etc. Different from the physical plans used in \cite{article}, such information has no unified form for encoding. We consider the conditions as text and encode using a pre-trained Sentence Transformers encoder \cite{reimers-2019-sentence-bert}. Such an encoder can capture the textual differences between conditions effectively and have unified embedding dimensions to simplify further analysis.
  \item \textbf{Cardinality and cost:} From \cite{qpeval} we observe that the estimated cardinality and cost are important in describing a query. We collect the row count and estimated cumulative cost values and normalise them through an MLP layer.
\end{itemize}
\vspace{-\topsep}
We simply concatenate the three information vectors together to be the encoded embedding for a node in the given query tree. We use the same tree Transformer model in \cite{article} to get the final representation of a query given its tree nodes’ embeddings. The final representation of the whole query will be computed by the tree-biased attention module.

\noindent \textbf{Contrastive learning structure.} 

Due to the necessity of executing queries, the volume of training triplets produced by our demonstration preparation module is limited. Unlike the query representation model in \cite{article}, which is trained directly on abundant labeled data, our approach requires a more sophisticated training framework to effectively capture query representation with the generated training data. Inspired by SimCSE \cite{gao2022simcse}, we design a contrastive learning structure to train our query representation model on the limited training data. In a training batch containing $N$ tuples, we consider each original query's improved query as its ``positive'' query, its degraded query as its ``hard negative'' query, and the remaining improved and degraded queries within the same batch as ``negative'' queries. This allows us to pull close distances between original queries and their improved versions while pushing apart those with degraded queries.
Following such setting, the loss $l_i$ for the $i_{th}$ tuple ($Q_i$, $Q_i^{+}$, $Q_i^{-}$) can be computed as
\begin{equation}
l_i = -\log\frac{e^{\textnormal{sim}(h_i, h_i^+)/\tau}}{\Sigma_{j=1}^N(e^{\textnormal{sim}(h_i, h_j^+)/\tau} + e^{\textnormal{sim}(h_i, h_j^-)/\tau})}
\label{equ:loss}
\end{equation}
where $\tau$ is a temperature hyper-parameter, $h_i$, $h_i^+$ and $h_i^-$ stand for the representation of $Q_i$, $Q_i^+$ and $Q_i^-$ respectively, and the function
$\textnormal{sim}(h_1, h_2)$ is the cosine similarity $\frac{h_1^Th_2}{\|h_1\|\cdot\|h_2\|}$. 

As an example in a training batch of size 2, for the first original query $Q_1$ shown in the bottom part of Figure \ref{fig:CS}, the positive query will be its corresponding improve query $Q_1^{+}$, and other in-batch improve or degrade queries $Q_1^{-}$, $Q_2^{+}$ and $Q_2^{-}$ are all regarded as negative queries. The final loss for the batch will be the sum of the losses for the two tuples.

\begin{figure}[t]
  \centering
  \includegraphics[width=\linewidth]{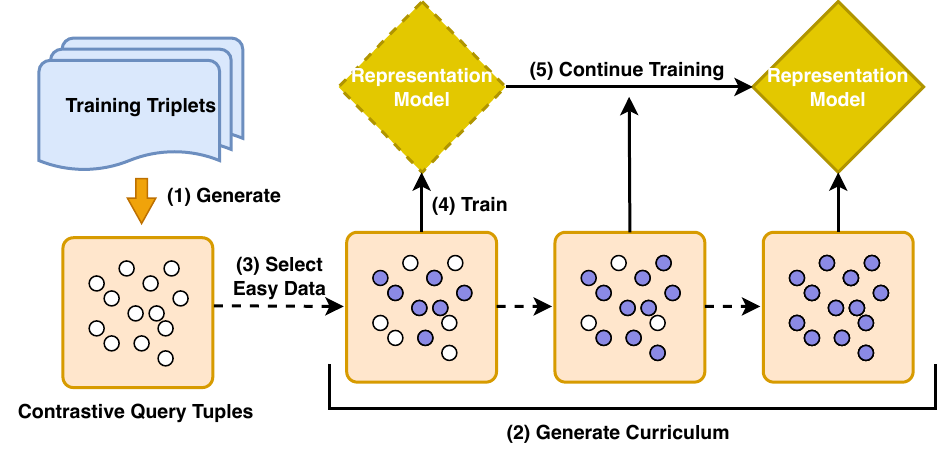}
  \vspace{-20pt}
  \caption{The overall curriculum learning pipeline to train the contrastive selector using generated training triplets.}
  \vspace{-10pt}
  \Description{the overall pipeline}
  \label{fig:pipeline}
\end{figure}

\subsection{Curriculum Learning Pipeline\label{sec:mod_lea}}
\label{sec:method_CL}

\textbf{Motivation.} Although we have developed a representation-based demonstration selector, training the contrastive model presents several challenges. First, unlike the original SimCSE approach used in natural language inference tasks, which benefits from abundant data~\cite{Cer_2017}, our model's training is constrained by data scarcity. Our contrastive query tuples, derived from a limited variety of training triplets, face scalability issues due to the high computational cost of query execution. Furthermore, the complexity of query representations in our model surpasses the simplicity of word embeddings used in SimCSE. Given these constraints—limited data and a complex training target—we propose adopting a curriculum learning pipeline. This approach is designed to enhance the learning efficiency and effectiveness of our contrastive representation model.

\begin{algorithm}[!b]
\caption{Contrastive Training under Curriculum Scheduler}\label{alg:curriculum}
\begin{algorithmic}[1]
\Require Total training data $S_0$, Number of iterations $I$
\Require Initialized model $E_0$
\State $N_0 = len(S_0)$ \Comment{Total number of training data }
\State $N = \lceil(N_0 / I)\rceil$ \Comment{Each iteration incremental data size }
\State $Tr_0 = \emptyset$ \Comment{Initial training data }
\State $i = 1$ \Comment{Initial iteration }
\While{$i \leq I$}
\If{$len(N) > len(S_{i-1})$} \Comment{If less than N data left}
    \State $Tr_i \gets S_{i-1}$ \Comment{Select all the data left}
    \State $Tr_i = Tr_{i-1} + Tr_i$  \Comment{Append to training data}
    \State Train $E_{i-1}$ on $Tr_i$ and get $E_i$ \Comment{Continue train the model}
\Else
    \State $C_i \gets Top_N(S_{i-1})$ based on $conf_{E_{i-1}}(\cdot)$ \Comment{Select N easy data following curriculum from the unvisited dataset}
    \State $S_i = S_{i-1} - C_i$  \Comment{Deduct them from unvisited data}
    \State $Tr_i = Tr_{i-1} + C_i$  \Comment{Append them to training data}
    \State Train $E_{i-1}$ on $Tr_i$ and get $E_i$ \Comment{Train the model}
    \State $i=i+1$ \Comment{Move to next iteration}
\EndIf
\EndWhile
\State Use the final $E_I$ for inference
\end{algorithmic}
\end{algorithm}

As depicted in Figure~\ref{fig:pipeline}, the essence of this pipeline is to strategically implement an effective curriculum. Starting with the provided training triplets, we initially train our contrastive representation model on a smaller, simpler subset, progressively incorporating easier subsets from the remaining dataset and retraining the model until all training data is utilized. The methodology for generating our curriculum is detailed in Algorithm \ref{alg:curriculum}. This algorithm begins with an empty model; each iteration involves selecting a subset of training data on which the current model performs with the highest confidence, followed by model retraining to incorporate this new subset (lines 5-17). This iterative retraining process continues until the entire training dataset has been incorporated.

In particular, we sample the easier subset of remaining training data by the confidence of the model to the data. Suppose we get the embeddings of two queries using our contrastive model to be $x$ and $y$, we can compute their similarity scores\ using the cosine similarity to keep consistency with the training objective in Equation \ref{equ:loss}. For each contrastive query tuple $\langle Q, Q^+, Q^- \rangle$, since we expect to have the $\textnormal{sim}(E(Q), E(Q^+))=1$ and $\textnormal{sim}(E(Q), E(Q^-)) = 0$, we define a confidence score of the contrastive model $E$ to a given tuple as: 
\begin{equation}
conf_{E}(Q) = \textnormal{sim}(E(Q), E(Q^+)) - \textnormal{sim}(E(Q), E(Q^-)) + 1
\end{equation}
Therefore, at each iteration $i$, given our trained model $E_{i-1}$, previous training dataset $T_{i-1}$ and the unvisited dataset $D_{i-1}$, we can generate the current tuples (denoted as $S_i$) with the highest confidence score in $D_{i-1}$. They are then moved into the training set, resulting in the new training set $T_i = T_{i-1} + S_i$ and the new unvisited dataset $D_i = D_{i-1} - S_i$.

\begin{table*}[!t]
\setlength\tabcolsep{2.5pt}
\begin{tabular}{lcccccccccccc}
\toprule
Execution time(sec)   & \multicolumn{4}{c}{TPC-H}       & \multicolumn{4}{c}{IMDB} & \multicolumn{4}{c}{DSB}   \\ \cmidrule(lr){1-1}\cmidrule(lr){2-5}\cmidrule(lr){6-9}\cmidrule(lr){10-13}
Method    & Mean  & Median & 75th  & 95th   & Mean & Median & 75th & 95th & Mean & Median & 75th & 95th \\ \hline \hline
\textbf{Original}  & 70.90 & 22.00  & 37.01 & 300.00 & 6.99 & 1.86   & 5.12 & 32.49  & 60.55   & 6.64   & 26.55 & 300.00 \\ 
\textbf{LR}        & 39.40 & 22.00  & 32.21 & \textbf{159.95} & 6.20 & 1.62   & 4.74 & 32.45 & 59.21 &  5.14 & 53.78 & 300.00  \\
\textbf{LLM only}        & 70.67 & 22.00 & 37.01 & 300.00 &  6.96 & 1.86   & 5.10 & 32.49 & 61.60 & 6.53 & 26.40 & 300.00 \\ \hline
\textbf{LLM-R\textsuperscript{2}} (Ours)   & \textbf{37.23} & \textbf{17.40}  & \textbf{29.80} & 164.12 & \textbf{3.91} & \textbf{1.33}   & \textbf{3.52} & \textbf{18.16} & \textbf{24.11} & \textbf{2.16}   & \textbf{12.61} & \textbf{196.61} \\ 
\% of \textbf{Original}   & 52.5\%  & 79.1\%   & 80.5\%  & 54.7\%   & 56.0\% & 71.3\%   & 68.7\% & 55.9\%  & 39.8\%  & 32.5\%   & 47.5\%  & 65.5\% \\ 
\% of \textbf{LR}  & 94.5\%  & 79.1\%   & 92.5\%  & 102.6\%  & 63.1\% & 82.0\%   & 74.3\% & 56.0\%  & 40.7\%  & 42.0\%   & 23.4\%  & 65.5\% \\ 
\% of \textbf{LLM only}  & 52.7\%  & 79.1\%   & 80.5\%  & 54.7\%   & 56.2\% & 71.3\%   & 69.0\% & 55.9\%  & 39.1\%  & 33.1\%   & 47.8\%  & 65.5\% \\ 
\bottomrule
\end{tabular}
\caption{Execution time v.s. different query rewrite methods}
\label{tab:main_results}
\vspace{-20pt}
\end{table*}

\section{Experiment}
In this section, we evaluate our proposed system's effectiveness, efficiency, and generalization capabilities. 

\subsection{Experimental Setup}
\subsubsection{Dataset}
We use three datasets from different domains for our evaluations:

\noindent \textbf{IMDB (JOB workload)~\cite{Leis2015HowGA}:} The IMDB~\cite{maas-EtAl:2011:ACL-HLT2011} dataset consists of data on movies, TV shows, and actors. It's utilized in conjunction with the Join Order Benchmark (JOB) to test a database management system's efficiency in executing complex join queries, and it comprises 5,000 queries.

\noindent \textbf{TPC-H~\cite{TPC-H.tools}:} A benchmark dataset for evaluating database management systems, generated using the official toolkit to include approximately 10 GB of data and 5,000 queries.

\noindent \textbf{Decision Support Benchmark (DSB)~\cite{ding2021dsb}:} This benchmark is developed to evaluate traditional database systems for modern decision support workloads. It is modified from the TPC-DS to include complex data distributions and challenging query templates, and it contains a total of 2,000 queries.

\subsubsection{Rewrite Rules} 
To enhance the efficiency of the rule proposal and rewriting process for subsequent experiments, we integrate Apache Calcite \cite{Begoli_2018} as our rewrite platform, alongside its comprehensive set of rewrite rules by following previous work \cite{10.14778/3485450.3485456}. Examples of utilized rewrite rules and their functions are illustrated in Table \ref{tab:rules}, with a complete enumeration available on the official website \cite{org.apache.calcite.rel.rules}. Specifically, we introduce a rule termed ``EMPTY'' to signify instances where the query remains unchanged, thereby standardizing LLM outputs with an indicator for scenarios that do not require query rewrite.

\subsubsection{LLM Setting}

We leverage the ChatGPT API~\cite{chatapi}, which is built upon the GPT-3.5-turbo architecture~\cite{brown2020language}. Furthermore, we assess our system's generalizability across other Large Language Models (e.g., GPT-4), as detailed in Section~\ref{exp:ablation}.


\subsubsection{Baseline Methods} We compare our system with two baseline methods:

\noindent \textbf{Learned Rewrite (LR)~\cite{10.14778/3485450.3485456}}: This approach, recognized as the state-of-the-art query rewrite method, incorporates a cost estimation model for predicting the performance of rewritten queries. It further employs a Monte Carlo Tree-based search algorithm to identify the optimal query.

\noindent \textbf{LLM only~\cite{li2023llm}}: This method straightforwardly generates a rewritten query from the input, incorporating task instructions, schema, and a fixed demonstration as prompts to the LLM. when the rewritten queries are not executable or equivalent to the original queries, we substitute them with the original queries. This ensures a fair comparison with rule-based methods.


\begin{table}[!t]
\setlength\tabcolsep{2.5pt}
\begin{tabular}{l|c|c|c}
\toprule
Counts   & \multicolumn{3}{c}{TPC-H/IMDB/DSB}       \\ \cmidrule(lr){1-1}\cmidrule(lr){2-4}
Method    & Rewrite \#  & Improve \# & Improve \%    \\ \hline \hline
\textbf{LR}        & 258/203/\textbf{456} & 192/197/193  & 74.42/\textbf{97.04}/42.32 \\
\textbf{LLM only}   & 197/102/210 & 68/67/8 & 34.5/65.68/3.81  \\\hline
\textbf{LLM-R\textsuperscript{2}} & \textbf{323}/\textbf{302}/341 &   \textbf{305}/\textbf{292}/\textbf{222}   &   \textbf{94.43}/96.69/\textbf{65.10} \\ 
\bottomrule
\end{tabular}
\caption{The rewritten queries' number v.s. different methods}
\label{tab:rewrite_cnt}
\vspace{-20pt}
\end{table}

\subsubsection{Training Setting.} 
In the demonstration preparation phase, we exclude any training queries already present in the demonstration pool from being selected as demonstrations to mitigate potential bias. For the development of our query representation-based demonstration selector, we adopt a curriculum learning strategy encompassing four iterations ($I=4$). Each iteration involves further training our contrastive representation model with a learning rate of $10^{-5}$, a batch size of $8$, over three epochs, utilizing a Tesla-V100-16GB GPU.

\subsubsection{Evaluation Metrics} 
For the evaluation of rewrite methods, two key metrics are employed: \textit{query execution time} and \textit{rewrite latency}, which are respectively employed to evaluate the executing efficiency and the computational efficiency. To mitigate variability, each query is executed five times on a 16GB CPU device, with the average execution time calculated after excluding the highest and lowest values. To address the challenge posed by overly complex queries that exceed practical execution times, a maximum time limit of $300$ seconds is imposed, with any query exceeding this duration assigned a default execution time of $300$ seconds. This approach facilitates a broader range of experimental conditions. For assessing rewrite latency—the time required to complete a query rewrite—a custom Python script is utilized to invoke both rewrite methods, capturing the average rewrite latency across all test queries on the same hardware platform.

\begin{table}[!t]
\begin{tabular}{l|c|c|c}
\hline
Total (Latency)        & TPC-H  & IMDB & DSB   \\ \hline \hline
\textbf{LR}	     & 40.98(1.58)   & 7.24(1.04)  & 60.99(1.78)\\ \hline
\textbf{LLM only}  & 75.37(4.70)   & 7.58(1.38)  & 64.21(6.00)\\ \hline
\textbf{LLM-R\textsuperscript{2}}  & \textbf{40.63}(3.40)   & \textbf{6.81}(2.90)  & \textbf{27.40}(3.29)\\ \hline
\end{tabular}
\caption{The rewrite performance in total average rewrite query execution time including the rewrite latency}
\label{tab:latency}
\vspace{-10pt}
\end{table}

\begin{table}[!t]
\begin{tabular}{l|c|c|c|c}
\hline
Execution time(sec)        & Mean  & Median & 75th  & 95th   \\ \hline \hline
Original  & 6.99 & 1.86   & 5.12 & 32.49  \\ \hline
\textbf{LLM-R\textsuperscript{2}}  & \textbf{4.41} & \textbf{1.35}   & \textbf{3.57} & \textbf{17.84} \\ \hline
\textbf{LR}  & - & -   & - & - \\ \hline
\textbf{LLM only}  & 6.99 & 1.86 & 5.12 & 32.49\\ \hline
\end{tabular}
\caption{Training on TPC-H and Testing on IMDB}
\label{tab:transferability}
\vspace{-25pt}
\end{table}

\begin{table*}[!t]
\setlength\tabcolsep{2.5pt}
\begin{tabular}{lcccccccccccc}
\toprule
Execution time(sec)   & \multicolumn{4}{c}{TPC-H 1G} & \multicolumn{4}{c}{TPC-H 5G}  & \multicolumn{4}{c}{TPC-H 10G}       \\ \cmidrule(lr){1-1}\cmidrule(lr){2-5}\cmidrule(lr){6-9}\cmidrule(lr){10-13}
Method    & Mean  & Median & 75th  & 95th  & Mean  & Median & 75th  & 95th   & Mean & Median & 75th & 95th \\ \hline \hline
Original  & 52.02 & 0.57  & 1.39 & 300.00 & 53.90&3.27&11.53&300.00 & 70.90 & 22.00  & 37.01 & 300.00  \\ 
\textbf{LLM-R\textsuperscript{2}}   & \textbf{15.19}  & \textbf{0.56}  & \textbf{1.14} & \textbf{55.20} & \textbf{19.34}&\textbf{3.20}&\textbf{7.97}&34.70 & \textbf{37.23}  & \textbf{17.40}  & \textbf{29.80} & 164.12  \\ 
\textbf{LR}   & 25.40  & 0.57 & \textbf{1.14} & 213.81  &  20.10&4.02&9.02&\textbf{32.14}  & 39.40  & 22.00  & 37.21 & \textbf{159.95} \\ 
\textbf{LLM only}   & 52.73&2.14&4.49&300.00 &  54.13&3.62&11.56&300.00  & 70.67  & 22.00  & 37.01 & 300.00 \\ 
\bottomrule
\end{tabular}
\caption{Execution time v.s. different data scales.}
\vspace{-20pt}
\label{tab:scales}
\end{table*}

\subsection{Executing Efficiency Evaluation}
As presented in Table~\ref{tab:main_results}, our study conducts a comparative analysis between our proposed method \textbf{LLM-R\textsuperscript{2}} and two baseline methods. We meticulously document the mean, median, 75th percentile, and 95th percentile values of execution times to provide a comprehensive performance evaluation. The mean and median offer insights into the general efficacy of the methods across the datasets, whereas the 75th and 95th percentiles facilitate an understanding of the methods' behavior for long tail cases. Our analysis yields several key observations:

\noindent \textbf{(1)} \textbf{LLM-R\textsuperscript{2}} demonstrates superior reduction of query execution time, outshining all baseline methods across the three datasets. Specifically on the TPC-H, IMDB and DSB datasets, \textbf{LLM-R\textsuperscript{2}} reduces the execution time of the queries on average to 94.5\%, 63.1\% and 40.7\% of the queries rewritten by baseline method \textbf{LR}, 52.7\%, 56.0\% and 33.1\% relative to \textbf{LLM only}, and even further 52.5\%, 56.0\% and 39.8\% compared to the original query. This performance enhancement is attributed to the optimization of demonstration selection for prompting the LLM-enhanced rewrite system, enabling \textbf{LLM-R\textsuperscript{2}} to suggest superior rewrite rules. Furthermore, leveraging an LLM-enhanced system, \textbf{LLM-R\textsuperscript{2}} offers more adaptable rule suggestions and better tailors these to the input query than does the \textbf{LR} baseline. 

\noindent \textbf{(2)} The margin of improvement over \textbf{LR} is notably greater in the IMDB and DSB datasets than in the TPC-H dataset. This discrepancy stems from two factors. First, TPC-H is also the mainly analysed dataset in \cite{10.14778/3485450.3485456} for \textbf{LR}. Most of the effective rewrite rules for TPC-H queries can already be applied by \textbf{LR}, leaving \textbf{LLM-R\textsuperscript{2}} with limited scope for further enhancements. Second, the TPC-H dataset's reliance on only 22 query templates results in a lack of query diversity, thus constraining the full demonstration of \textbf{LLM-R\textsuperscript{2}}'s superiority utilizing LLM generalisation and reasoning abilities.

\noindent \textbf{(3)} \textbf{LR}'s under-performance in the DSB dataset can be attributed to its design limitations adopting a greedy search algorithm. The DSB dataset, being entirely new and unmet for \textbf{LR}, poses unique challenges. Moreover, the Monte Carlo tree search algorithm employed by \textbf{LR}, with its greedy search strategy that retains only a select few best options at each step, struggles with the dataset's complex and expensive query trees. This limitation makes it difficult for the algorithm to select the most effective rules, to which explains its poor performance in handling the DSB dataset's demands.

\noindent \textbf{(4)} \textbf{LLM only} has the worst performance. We observe that LLMs struggle to effectively address the query rewrite challenge, and has only marginal reductions in mean cost on the TPC-H dataset and median cost on the DSB dataset. Given that non-executable or non-equivalent rewrite attempts are categorized as 'no rewrite,' many rewritten queries are the same as the original queries across the datasets.

Furthermore, we evaluate the performance by collecting statistics on the number of successful rewrites performed by each method across three datasets. As shown in Table \ref{tab:rewrite_cnt}, we observe that:

\noindent \textbf{(1)}  \textbf{LLM-R\textsuperscript{2}} excels by having the most efficiency-enhancing rewrites, achieving the largest improvement percentage upon rewriting. Compared to the baseline, \textbf{LLM-R\textsuperscript{2}} has both a higher number of rewrites and a significant improvement in query execution efficiency across all the evaluated datasets.

\noindent \textbf{(2)} \textbf{LLM only} fails in most of its rewrite attempts. We look into the rewrites which do not return the same results as the original queries in the TPC-H dataset, 119 of the total 129 queries are either not consistent with the original one or have errors to execute. Similarly, 193 of the 202 attempts to rewrite failed in the DSB dataset, since the DSB queries and schema would be too complicated for the LLM. This observation aligns with the results in \cite{li2023llm}, in which the text-to-SQL task only achieved around 40\% accuracy with carefully designed prompts. Although the IMDB dataset is simpler compared to TPC-H and DSB datasets, where \textbf{LLM only} only fails 31 of the total 102 attempts, the LLM makes limited effective rewrites due to lack of database and query structure knowledge. In contrast, our \textbf{LLM-R\textsuperscript{2}}, which benefits from both the reasoning ability of LLM and the rewrite ability of database platforms, is able to rewrite more queries successfully and have as higher rewrite improvement rate across all the datasets.

\subsection{Computational Efficiency Evaluation}
To evaluate the computational efficiency, we rigorously assess the average rewrite latency for input queries across all datasets for the \textbf{LLM-R\textsuperscript{2}} framework as well as the \textbf{LR} and \textbf{LLM only} baselines. Moreover, to ascertain if query time reduction adequately compensates for the rewriting latency, we combine the execution cost and rewrite latency to formulate a comprehensive metric. As delineated in Table~\ref{tab:latency}, our analysis yields significant insights:

\noindent \textbf{(1)} \textbf{LLM-R\textsuperscript{2}} incurs additional latency compared to \textbf{LR}, specifically requiring an average of 1.82, 1.86, and 1.51 seconds more to rewrite queries from the TPC-H, IMDB, and DSB datasets, respectively. This heightened latency is due to our system's complexity. Notably, \textbf{LLM-R\textsuperscript{2}} employs a demonstration selection model and leverages the online LLM API, which together account for the increased rewrite latency.

\noindent \textbf{(2)} However, the increased rewrite latency in our system \textbf{LLM-R\textsuperscript{2}} is justifiable given that the sum of rewrite latency and execution time is lower than that of baseline methods, especially for the most complicated DSB queries. This indicates that the complex queries benefit more from our method.

\noindent \textbf{(3)} The \textbf{LLM only} approach exhibits considerable latency as the LLM endeavors to directly generate a rewritten query, underscoring the complexity of direct SQL query generation for LLMs. This latency becomes more pronounced with the complexity of the query and database, notably in the TPC-H and DSB datasets. The comparison between our \textbf{LLM-R\textsuperscript{2}} framework and the \textbf{LLM only} approach demonstrates that our methodology, which focuses on generating rewrite rules, is more effectively processed by LLMs.

\begin{table*}
\setlength\tabcolsep{2.5pt}
\begin{tabular}{lcccccccccccc}
\toprule
Execution time(sec)   & \multicolumn{4}{c}{TPC-H}       & \multicolumn{4}{c}{IMDB} & \multicolumn{4}{c}{DSB}   \\ \cmidrule(lr){1-1}\cmidrule(lr){2-5}\cmidrule(lr){6-9}\cmidrule(lr){10-13}
Method    & Mean  & Median & 75th  & 95th   & Mean & Median & 75th & 95th & Mean & Median & 75th & 95th \\ \hline \hline
Original  & 70.90 & 22.00  & 37.01 & 300.00 & 6.99 & 1.86   & 5.12 & 32.49  & 60.55   & 6.64   & 26.55 & 300.00 \\ \hline
\textbf{Zero-shot} & 46.15 & 21.95  & 33.26 & 300.00 & 6.98 & 1.85 & 5.12  & 32.49  & 34.53 & 3.35 & \textbf{11.52} & 300.00  \\ 
\textbf{Random}    & 40.50 & 21.63  & 32.22 & 165.63 & 5.45 & 1.70   & 4.50 & 25.03 & 45.88 & 5.43 & 17.41 & 300.00 \\ 
\textbf{Tree}      & 39.21 & 18.97  & 30.89 & \textbf{164.10} & 4.40 & \textbf{1.24}   & \textbf{3.40} & 18.89 & 26.10 & 3.86 & 13.54 & 240.74 \\ 
\textbf{SentTrans} & 40.19 & 19.21  & 32.21 & 164.99 & 6.05 & 1.70   & 4.49 & 30.01 & 24.68 & 3.95 & 13.18 & 197.23 \\ \hline
\textbf{LLM-R\textsuperscript{2}}  & \textbf{37.23} & \textbf{17.40}  & \textbf{29.80} & 164.12 & \textbf{3.91} & 1.33   & 3.52 & \textbf{18.16} & \textbf{24.11} & \textbf{2.16}   & 12.61 & \textbf{196.61} \\ 
\bottomrule
\end{tabular}
\caption{Execution time v.s. different selection approaches.}
\vspace{-20pt}
\label{tab:ablat_main_results}
\end{table*}

\subsection{Robustness Evaluation}
We next evaluate the robustness of our \textbf{LLM-R\textsuperscript{2}} framework, focusing on two critical dimensions: transferability and flexibility. Transferability evaluates the system's ability to generalize across diverse datasets, while flexibility examines whether \textbf{LLM-R\textsuperscript{2}} maintains its high performance as the volume of data increases. These aspects are crucial for understanding the adaptability and efficiency of \textbf{LLM-R\textsuperscript{2}} in varied environments.

\subsubsection{Transferability across different datasets}
In order to evaluate our method's transferability, we used the demonstration selection model trained on the TPC-H dataset to rewrite queries in the IMDB dataset. As shown in Table~\ref{tab:transferability}, the results reveal our method's transferred performance is comparable with the in-distribution trained method and highly superior over \textbf{LLM only} when applied to a different dataset. \textbf{LLM only} fails to make effective rewrites given the fixed demonstration from the TPC-H dataset, where most rewrites lead to meaningless changes like removing table alias. Since \textbf{LR}'s cost model lacks cross-dataset transfer capability, its results are not available. These findings suggest the potential to develop a robust model by combining multiple datasets, enhancing its ability to address a wide array of unseen queries and datasets.

\subsubsection{Flexibility across different data scales}
To further analyse the flexibility of our method, we regenerate the TPC-H dataset using different scale factors. We additionally generate TPC-H dataset using scale factor 1 (around 1GB data) and 5 (around 5GB data) apart from 10 in the main results to simulate a change of database size. From scale factor 1 to 10, we can see in Table \ref{tab:scales} the efficiency of queries rewritten by our method increases consistently and surpasses the baseline methods.

\subsection{Ablation Studies\label{exp:ablation}}

We conduct an ablation study to evaluate our method's performance along two distinct dimensions: \textit{different selection approaches} and \textit{specific settings in the selection model}. At first, we explore alternative selection approaches by substituting the learned selection model with different approaches to gauge their impact. Subsequently, we delve into the intricacies of the selection model by replacing individual components of the model.

\subsubsection{Different selection approaches} 
We design the following approaches to replace the contrastive selection model in our system: 


- \textbf{Zero-shot:} This method employs the LLM-R\textsuperscript{2} to rewrite input queries without any preliminary demonstrations.

- \textbf{Few-shots:} Building on insights from Section~\ref{sec:method_DM}, we refine the demonstration pool with three intuitive methods for one-shot demonstration selection: \textbf{Random}, \textbf{Tree}, and \textbf{SentTrans}. 

Table 7 shows the results and we make the following observations:

\noindent \textbf{(1) Effectiveness of the LLM-enhanced system:} The \textbf{Zero-shot} approach outperforms the original queries significantly, which indicates that the LLM-R\textsuperscript{2} component within our rewrite system is capable of enhancing original queries, showcasing the underlying potential of the LLM to offer viable query rewrite suggestions. This observation suggests that even though the recommendations provided may not always be optimal—owing to constraints such as incomplete information and occasional inaccuracies—the LLM's contributions are valuable in improving query performance. 

\noindent \textbf{(2) Effectiveness of introducing demonstrations:}  We observe that approaches incorporating demonstrations into the rewrite system consistently surpass the Zero-shot setting across all datasets. The sole exception is observed with the \textbf{Random} method, which falls short of the Zero-shot rewrite performance on the DSB dataset. This observation underscores the significance of leveraging demonstrations to enhance the rewrite system, significantly boosting the quality of rewrites. Furthermore, the improvement across diverse datasets highlights the universal applicability and effectiveness of demonstration-based prompting in refining rewrite outcomes. 

\noindent \textbf{(3) Effectiveness of the contrastive selection model:} Our comparative analysis underscores the significance of selecting high-quality demonstrations for query rewriting. The findings reveal that superior demonstrations directly contribute to the generation of more effective rewritten queries.

\begin{table}[]
\begin{tabular}{l|c|c|c|c}
\hline
Execution time(sec)        & Mean  & Median & 75th  & 95th   \\ \hline \hline
Original  & 70.90 & 22.00  & 37.01 & 300.00 \\ \hline
\textbf{LLM-R\textsuperscript{2}} (1-shot)  & \textbf{37.23} & \textbf{17.40}  & \textbf{29.80} & \textbf{164.12} \\ \hline
\hspace{0.2cm} \textbf{w/o Curriculum} & 38.73 & 19.70  & 32.17 & 164.98   \\ \hline
\textbf{LLM-R\textsuperscript{2}} (3-shots)  & 54.08 & 19.67  & 37.01  & 300.00 \\ \hline
\textbf{LLM-R\textsuperscript{2}} (GPT-4)  & 38.58 & 20.32 & 32.27 & 167.26 \\ \hline
\end{tabular}
\caption{Performance comparison of LLM-R\textsuperscript{2} with and without the curriculum learning pipeline on TPC-H}
\vspace{-25pt}
\label{tab:curriculum}
\end{table}

\begin{figure*}[!t]
  \centering
  \includegraphics[width=0.9\linewidth]{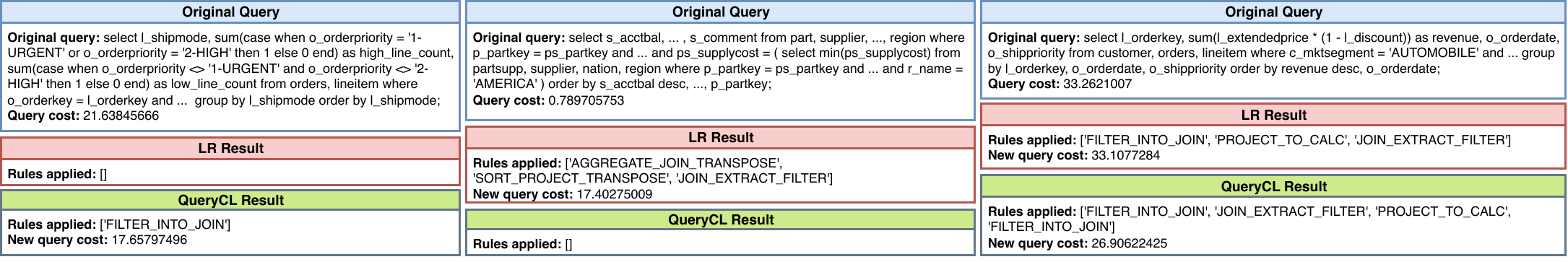}
  \vspace{-10pt}
  \caption{Examples of the rewrite results of baseline Learned Rewrite method and out LLM-R\textsuperscript{2} method.}
  \Description{rewrite_examples}
  \vspace{-10pt}
  \label{fig:rewrite_examples}
\end{figure*}

\subsubsection{Effectiveness of specific settings in the selection model.} In this experiment, we concentrate on assessing three critical aspects within the contrastive selection model:

- \textbf{The Curriculum Learning pipeline:} We investigate the curriculum learning pipeline's efficacy by comparing it with a baseline model. Specifically, this baseline involves training a selection model on the TPC-H dataset using all training triplets simultaneously, rather than employing a curriculum learning-based approach.

- \textbf{Demonstration Quantity:} We evaluate the impact of varying the number of demonstrations by focusing on the most prevalent configurations—namely, 1-shot and 3-shot demonstrations. This experiment aims to elucidate the demonstration quantity's effect on the model's performance.

- \textbf{Different LLMs:} We explore the implications of integrating GPT-4, a more advanced LLM recognized for its superior capabilities in natural language processing, into our rewriting system. Given the financial implications of utilizing the GPT-4 API, our experimental setup restricts the use of GPT-4 to the enhancement of the test dataset rewrite process, with demonstrations and models derived from GPT-3.5-turbo.

Table \ref{tab:curriculum} shows the evaluation results and we obtain the following key insights:


\noindent \textbf{(1)} Our query representation model demonstrates superior performance in selecting optimal demonstrations compared to baseline approaches, and the incorporation of a curriculum-based training methodology significantly amplifies this advantage. For instance, direct training on the complete dataset results in a notable reduction in execution cost, averaging a decrease of 32.17 seconds and a median of 2.3 seconds, respectively. Utilizing the curriculum learning approach for training the demonstration selector further contributes to cost efficiency, achieving an average reduction of 1.5 seconds and a median decrease of 2.3 seconds. These findings underscore the efficacy of our proposed query representation model and the curriculum learning framework. 

\noindent \textbf{(2)} Employing a 3-shot approach, as opposed to a 1-shot strategy, adversely affects performance. A detailed examination of the rewritten queries reveals that, the 3-shot method generated only 255 rewrite proposals, and 235 of these rewrites yielded improvements in query execution efficiency. Despite a high success rate of 92.16\% for these rewrites, the primary limitation lies in the significantly reduced number of rewrite suggestions. This reduction is largely attributed to the inconsistent guidance provided by the three demonstrations. Additionally, the increased cost of rewrites and the challenges posed by longer in-context texts for LLM analysis emerge as critical yet unresolved issues when employing 3-shot prompting. Based on these findings, we deduce that 1-shot prompting presents a more efficient and effective approach under the current experimental conditions.

\noindent \textbf{(3)} Despite GPT-4's enhanced capabilities, transitioning to a different model for inference adversely impacts the efficacy of our method. This observation underscores the complexity of optimizing performance within our proposed framework and suggests that consistency in model usage throughout the process may be pivotal for achieving optimal selection.

\subsection{Qualitative Analysis}

we proceed to present examples to illustrate the rewrite quality between various methods, focusing particularly on comparisons between our approach and baseline methods. Notably, due to the high incidence of erroneous rewrites generated by the \textbf{LLM-only} method, our analysis primarily compares our method against the \textbf{LR} baseline. Figure~\ref{fig:rewrite_examples} demonstrates our findings demonstrate the superior robustness and flexibility of our model compared to \textbf{LR}. For instance, in the first case study, our \textbf{LLM-R\textsuperscript{2}} method uncovers rewrite rules that remain undetected by \textbf{LR}. This discrepancy can be attributed to \textbf{LR}'s potentially ineffective cost model, which might erroneously consider the original query as already optimized. Conversely, our LLM-enhanced system suggests a rewrite that evidences significant potential for cost reduction. In the second case, \textbf{LR} is observed to occasionally transform an efficient query into a less efficient one. In the third scenario, \textbf{LLM-R\textsuperscript{2}} outperforms by modifying the rule sequence and incorporating an additional ``FILTER\_INTO\_JOIN'' operation, transforming a ``WHERE'' clause into an ``INNER JOIN'', thereby achieving a more efficient query rewrite than that offered by \textbf{LR}.

\begin{table}[!t]
\setlength\tabcolsep{2.5pt}
\begin{tabular}{lcccccc}
\toprule
Counts   & \multicolumn{2}{c}{TPC-H}       & \multicolumn{2}{c}{IMDB} & \multicolumn{2}{c}{DSB}   \\ \cmidrule(lr){1-1}\cmidrule(lr){2-3}\cmidrule(lr){4-5}\cmidrule(lr){6-7}
Method    & Unique  & Total & Unique  & Total & Unique  & Total \\ \hline \hline
LR        & 5 & 405  & 1 & 192 &  9 & 707 \\ \hline 
LLM-R\textsuperscript{2}   & \textbf{56} & \textbf{1824}  & \textbf{6} & \textbf{361} &  \textbf{37} & \textbf{920}\\ 
\bottomrule
\end{tabular}
\caption{The variety of rules applied by the methods in terms of unique rules and total applications.}
\label{tab:rule_variety}
\vspace{-25pt}
\end{table}

Furthermore, we delve into the diversity of rewrite rules suggested by the different methods. Here, the term \textit{Unique} refers to the distinct categories of rewrite rules recommended by a method, whereas \textit{Total} denotes the aggregate count of all rewrite rule instances proposed. As illustrated in Table~\ref{tab:rule_variety}, it is evident that \textbf{LLM-R\textsuperscript{2}} not only recommends a higher quantity of rewrite rules but also exhibits a broader spectrum of rewrite strategies by employing a diverse range of rules. This observation underscores \textbf{LLM-R\textsuperscript{2}}'s enhanced flexibility and robustness, showcasing its capability to generate more varied and effective rewrite plans.

\section{Conclusion}

Despite the analysis above, we would like to point out the current limitation for further work. The main limitation for our LLM-R\textsuperscript{2} lies in the higher rewrite latency compared to DB only methods. Compared to traditional DB methods, calling LLM API and selecting demonstrations indeed consume more time. However, as shown in the experiment results, such higher latency can be alleviated by the larger execution time LLM-R\textsuperscript{2} decreases, and there is no doubt that our LLM-R\textsuperscript{2} is a successful example of exploring the LLMs’ application in database problems. We believe that the strong generalisation and reasoning ability of the LLMs can also be applied to other important database problems as well. In addition, further work can also be made to improve our current LLM enhanced query rewrite system, for example, utilising efficient demonstration selection algorithms like Faiss \cite{douze2024faiss}, or even specially fine-tune a LLM on query rewrite with more dataset.

To conclude, we propose a LLM-enhanced query rewrite pipeline to perform efficient query rewrite. By collecting useful demonstrations and learning a contrastive demonstration selector to modify the rewrite system inputs, we are able to successfully improve the input queries' efficiency across popular datasets. In addition, we further prove the effectiveness of our learning pipeline and the transferability of our method over different scales, model backbones and datasets, showing that LLM-enhanced methods could be an effective solution for efficiency-oriented query rewrite.

\bibliographystyle{ACM-Reference-Format}
\bibliography{sample}

\end{document}